\documentclass[sigconf]{acmart}

\acmConference[ICSE 2024]{46th International Conference on Software Engineering}{April 2024}{Lisbon, Portugal}

\copyrightyear{2024}
\acmYear{2024}
\setcopyright{acmlicensed}\acmConference[CHASE '24]{2024 IEEE/ACM 17th International Conference on Cooperative and Human Aspects of Software Engineering }{April 14--15, 2024}{Lisbon, Portugal}
\acmBooktitle{2024 IEEE/ACM 17th International Conference on Cooperative and Human Aspects of Software Engineering (CHASE '24), April 14--15, 2024, Lisbon, Portugal}
\acmDOI{10.1145/3641822.3641867}
\acmISBN{979-8-4007-0533-5/24/04}

\usepackage{array}
\usepackage{color,soul}
\usepackage{xcolor}

\title{Learning From Lessons Learned: Preliminary Findings From a Study of Learning From Failure}

\author{Jonathan Sillito}
\email{sillito@byu.edu}
\affiliation{%
  \institution{Brigham Young University}
  \city{Provo}
  \state{UT}
  \postcode{84602}
  \country{USA}
}

\author{Matt Pope}
\email{mattpope@byu.edu}
\affiliation{%
  \institution{Brigham Young University}
  \city{Provo}
  \state{UT}
  \postcode{84602}
  \country{USA}
}

\begin{document}

\begin{abstract}

Due to various sources of uncertainty, emergent behavior, and ongoing changes, the reliability of many socio-technical systems depends on an iterative and collaborative process in which organizations (1) analyze and learn from system failures, and then (2) co-evolve both the technical and human parts of their systems based on what they learn. 
Many organizations have defined processes for learning from failure, often involving postmortem analyses conducted after any system failures that are judged to be sufficiently severe. Despite established processes and tool support, our preliminary research, and professional experience, suggest that it is not straightforward to take what was learned from a failure and successfully improve the reliability of the socio-technical system. 
To better understand this collaborative process and the associated challenges, we are conducting a study of how teams learn from failure. We are gathering incident reports from multiple organizations and conducting interviews with engineers and managers with relevant experience. Our analytic interest is in what is learned by teams as they reflect on failures, the learning processes involved, and how they use what is learned.
Our data collection and analysis are not yet complete, but we have so far analyzed 13 incident reports and seven interviews. In this short paper we (1) present our preliminary findings, and (2) outline our broader research plans.

\end{abstract}

\maketitle

\section{Introduction}

In the winter of 2018, during a team lunch a group of software engineers from a large technology company discussed a recent system failure that had affected millions of users. The team had already completed a formal analysis of the incident and had written an incident report that captured the contributing factors, the lessons learned and the steps they planned to take to avoid a repeat of the incident. The trigger for the incident was a configuration change that had unintended consequences and so in the report the engineers summarized a key lesson as \emph{configuration is code}. During the lunch, the engineers debated how the lesson should be interpreted and what its implications were for the organization. 
Everyone was frustrated that configuration changes frequently triggered incidents; while specific issues had been fixed, it was not clear that the lesson was in fact being learned by the organization in any meaningful sense. 

This lunchtime experience, and others like it, are the motivation for this research. Naturally there are many types of failures that an organization may learn from (late delivery, for example) but in this research we are only interested the learning from what we term \emph{incidents}, which are software system failures that are considered serious enough that a postmortem is conducted, and an incident report published by the system owners. And so the terms \emph{failure} and \emph{incident} are generally used interchangeably in this paper.

Previous research considers how to effectively conduct postmortems (e.g.,~\cite{bjornson2009improving,dalal2013empirical}) and proposes associated process models~\cite{kohn2013integrated} and a vibrant practitioner communities encourages various best practices.\footnote{See https://learningfromincidents.io, for example} However, our preliminary research, and professional experience, suggest that there are still many challenges around how to take what was learned from a failure and improve the reliability of the socio-technical~\cite{davis2014advancing} system. And so the goal of our ongoing research is to deepen our understanding of this learning process (both the formal and informal aspects) and build an organizational learning model of the collaborative process that captures insights about how learnings and proposed changes are managed over time (see Section~\ref{discussion} for more details on our goals). To this end, we are gathering and analyzing data about the learning process used by organizations. So far the data has included publicly published incident reports and interviews with experienced practitioners. The goal of this short paper is to report on some of our preliminary findings, and outline some of our broader research plans.

\section{Background}

In this work, we expect to build on previous research in a number of (partially overlapping) areas including organizational learning, learning from failure, resilience engineering, socio-technical systems, cognitive bias, and others. Our research methodology is largely inductive and at this stage it is not clear how well that previous research will fit our data, however, we describe some previous work below at it represents sensitizing concepts that may partially guide our analysis~\cite{bowen2006grounded} or lenses through which we can view the data. Part of our work will be to assess how well various existing frameworks capture the important insights in our data.

\subsection{Organizational Learning from Experience} 


Organizational learning theory refers to learning as a change within an organization that occurs as a function of experience in efforts to achieve some organizational goal~\cite{argote2011organizational}. Learning by an organization involves changes in routines, systems, processes and social norms that determine behavior--beyond what is stored in any one person's head~\cite{smerek2017organizational}. The relevant theories explore the underlying processes involved and the relationship between \emph{experience} and \emph{change}. Organizational learning can be described as four learning processes (knowledge search, knowledge creation, knowledge retention and knowledge transfer) and we are interested in how these processes capture what is learned from incidents. 


\subsection{Learning From Failure} 

One kind of experience that organizations can learn from is related to \emph{failures}. 
The way individuals, teams and organizations react to and learn from failure has been studied in multiple domains: economics, organizational theory, strategy, sociology, psychology, etc. But generally, organizational learning from \emph{failure} considers failures and their (distinct) effects on organization routines, processes, strategies, and outcomes~\cite{desai2017organizational}. 
Presumably a learning that does not result in such a change may not have been learned, or it may have been learned by some individuals but not the organization. 

In the organizational learning context, failure may be defined as events, experiences or other occurrences that fail to meet expectations. In a software system context, a failure occurs when the service being delivered does not comply with the specification, or perhaps more informally does not meet the expectations various stakeholders have for the system~\cite{laprie1992dependability}. Expectations and organizational goals (and therefore the nature of the failures that attract attention) may shift overtime~\cite{pache2013inside}. For our analysis, the decision to treat an event as a failure is naturally up to the organization that experiences the failure. Longer term we are interested in exploring a broader range of experience (near misses~\cite{eloff2017software}, hypothetical experiences), but we will ignore those in this paper.

The reaction of an organization to failure has been shown to depend on various factors including goals and aspirations, strategies and market position, amount of conservable organizational slack~\cite{iyer2008performance}. Learning at the organizational level involves changing attributes of the organization that may resist change, based on direct or indirect experience and how individuals interpret that experience. Failures (as interpreted by the organization) play an important role, though \emph{attention} may drift toward or away from experiences. Of course, learning from failure is not an organization's only goal and relative focus on innovation versus safety mediates efforts around learning~\cite{haunschild2015organizational}.

\subsection{Resilience engineering} 

The field of \emph{resilience engineering} considers both failures and successes as possible outcomes of normal performance variability and that safety cannot be achieved by constraining or eliminating that~\cite{hollnagel2006resilience}. It follows then that it is necessary to study both successes and failures, and to find ways to reinforce the variability that leads to successes as well as dampen the variability that leads to adverse outcomes. Processes that enable an effective response to unexpected events and vulnerabilities that lie outside the scope of formal procedures can be described as resilience. Four organizational abilities contribute to such efforts: (1) responding to disturbances, (2) flexible monitoring, (3) anticipating disruptions, and (4) learning from experience. Strengthening those abilities is at the heart of resilience engineering's take on increasing resilience of socio-technical systems~\cite{furniss2011resilience}. In this context, lessons learned from failure may be captured as changes to procedures, changes to roles and functions, or changes to the organization itself.

Resilience engineering has been considered in a software context in various ways, including analyzing outages in internet-facing systems~\cite{grayson2019cognitive}, error handling~\cite{lopez2020bumps} and software development practice at the team level~\cite{lopez2023accounting}. In this work we will consider how the principles of resilience engineering, including many not discussed above, may contribute to our analysis of learning from failure.



\section{Methodology}

Our analytic interest is in what is learned by teams as they reflect on failures, the learning processes involved, and also how they use what is learned. To this end we are gathering incident reports from multiple organizations and conducting interviews with engineers and managers that are involved in relevant activities. We are taking an iterative, inductive, and deeply contextual analytic approach that will ground our results in data about real systems and the experiences of engineers and organizations. So far, we have analyzed 13 incident reports (with 50 lessons learned) and seven interviews. 

\subsection{Incident Report Collection and Analysis}

\begin{table*}
    \centering
    \caption{Examples lessons learned. The first column is an identifier for the incident, followed by the context or source of the learning, the lesson learned, and the number of actions planned.}\vspace{-0.8em}
    \label{tbl:examples}
    \begin{tabular}{l
        >{\raggedright\arraybackslash}p{2.9in}
        >{\raggedright\arraybackslash}p{3.3in}
        r}

        \toprule
        \textbf{ID} & \textbf{Incident context} & \textbf{Lesson learned} & \textbf{\#} \\

        \midrule

        I1 &
        Downgrading a database led to a failure once daily peak load was reached \emph{(operator actions)} &
        Always load test major infrastructure changes before deployment to production \emph{(testing practices)} &
        1 \vspace{0.25em}\\

        I7 & 
        User growth caused excessive load and failures across multiple systems \emph{(triggered by load)}&
        Small inefficiencies are significant at scale when aggregated across multiple systems \emph{(performance \& efficiency)} &
        4 \vspace{0.25em}\\

        I10 &
        Automation failed and responders struggled to restore service manually \emph{(automation \& support systems)}&
        Automation of operational tasks erodes human knowledge and leads to reliance \emph{(skill development)} &
        1 \vspace{0.25em}\\

        I12 &
        Failure cascaded across multiple systems, starting with single change \emph{(cascading effects)}&
        Shared cluster was single point of failure due to data partitioning strategy \emph{(failure isolation)} &
        1 \vspace{0.25em}\\


        \bottomrule
    \end{tabular}\vspace{-1em}
\end{table*}

Incident reports are the product of internal postmortem analyses and generally describe significant failures along with explanations of how they occurred, and how they were mitigated. Importantly for our research, they often also include descriptions of what was learned and what changes are planned to prevent future failures. To capture the variation inherent in this space, we intend to include many publicly available incident reports from many organizations. We are sampling from several public repositories of incident reports, most notably the Verica Open Incident Database\footnote{https://www.thevoid.community} because of its size and the metadata that supports various ways to categorize reports. Incident reports tend to represent a careful analysis by experienced engineers and managers, and we have found them to be a rich dataset and suitable for a grounded theory style analysis~\cite{corbin2008techniques}. To date we have analyzed 13 incident reports (from companies such as GitHub, Cloudflare, Slack and Epic Games), and in this paper we refer to those incident reports as I1\ldots I13.\footnote{Extracted data is available here: https://shorturl.at/lmrAB}

In reviewing each incident report, we are interested in any information about the organization's learning process. We are also (more concretely) extracting each of the explicitly listed lessons learned. These are not what we as analysts think they should have learned or could have learned, but what the authors of the incident reports have clearly stated that they learned. For each lesson learned, we also extract elements of the incident associated with that lesson (such as what led to the lesson), and list actions that connect with those lessons learned. These actions capture what is expected to happen after the postmortem with what was learned. Table~\ref{tbl:examples} summarizes several examples of our extracted data.

\subsection{Conducting and Analyzing Interviews}

To situate learning in the larger organizational context, we are also conducting a series of interviews using a protocol that has been approved by the Institutional Review Board at Brigham Young University. All our interview participants have experience with learning from failure, and we have intentionally recruited participants from organizations that vary along key dimensions (e.g., size and domain) and that play different roles in their organizations. So far, we have interviewed engineers, managers, and product managers, and in this paper we refer to them as P1\ldots P7. 

The interviews are open-ended and begin with a general prompt such as: ``we are interested in how you learn from incidents; we would love to hear anything you can tell us.'' Through the interviews we ask the participants follow-up questions about what was learned, how it was learned and what happened with what was learned over time. During the interviews we encourage the participants to talk about specific incidents and associated lessons rather than talk more generally. We are still conducting interviews, and we are still early in our analysis of the interview data collected so far, but we will present our preliminary findings that we find most interesting. 

\section{Preliminary Findings}

We discuss our preliminary findings in four parts. First, we discuss the events or aspects of an incident that motivate the lessons being learned. Next, we discuss what is being learned. Then we discuss the actions that are planned in response to lessons learned. Finally, we discuss important concepts for understanding the formal and informal learning processes. The first three subsections are primarily from our analysis of incident reports. The last subsection is primarily (though not exclusively) based on our analysis of the interviews. We are working on a more integrated analysis now. 

\subsection{What Parts of Incidents are Learned About?}

Incidents, as described in incident reports, have a life that may begin before the initial failure is triggered and continue at least until the failures are resolved. Various internal and external events, actions by various responders, and the behavior of primary and support systems are described. In our analysis we are identifying the different events in an incident's lifecycle that motivated each lesson learned in our data set in order to provide additional context for the learning process.
So far in our preliminary analysis we have identified five categories of events that led to lessons learned, including: (1) the triggering of incidents, or how they began, (2) events around responders finding out about an incident or determining the nature of the incident, (3) the ways an initial failure cascades and the scope of its impact, (4) the consequences of responder actions, and (5) how automation and support systems influence the progress of the incident. Interestingly, some sources of learning reflect actions or activities that ``worked'' or had the intended effect but were still seen as a learning opportunity. Similarly, responders also learned from problems that did not occur, but on reflection they realized could have occurred.

Due to space limitations, we will briefly share a few details about just two of these categories, starting with the triggering of incidents. Even if there is a clear trigger for an incident (say the deployment of a defective configuration), we have found that the multiple factors, events, and actions that led to the incident are potential sources of learning. So far these have included (note that the number of relevant lessons in our dataset is shown in brackets):

\begin{enumerate}
    \item The way that operator actions (such as maintenance activities) contributed to failures being triggered (7),
    \item The role of latent defects and the situations under which they are eventually triggered (2),
    \item How coincident events or conditions combined to contribute to a failure (2), and
    \item The consequences of growth and increased load throughout a system (5).
\end{enumerate}

The detection and diagnosis of a failure was difficult at times, and so was the source of multiple lessons. For example, during incident I2, ``despite several engineers reviewing the configuration changes'' that triggered the incident, everyone ``dismissed them as irrelevant''. Aspects of detection and diagnosis that proved to be learning opportunities included:


\begin{enumerate}
    \item Notification system failures that were correlated with or coincidental to the main system failure (3), 
    \item Ways in which monitoring, and the available metrics and tools were of limited use (4), and
    \item Cognitive limits, misunderstandings, etc (2).
\end{enumerate}





In summary, postmortem analysis covers more than just identifying a single root cause of an incident. The whole life cycle of the incident leads to lessons, which is a testament to the complexity that can be involved in operating and supporting such software systems and demonstrates that incidents can be a rich source of learning.  


\subsection{What is Being Learned?}

The lessons learned we are identifying can be categorized along multiple dimensions, and our categorizations are still evolving, but we have seen that one interesting way to organize the lessons is based on the parts of the larger socio-technical system being learned about. So far in our preliminary analysis, we have divided this into six categories: (1) core systems operated by the organization, (2) incident response practices and processes, (3) monitoring, logging, and alarming systems, (4) other support systems and infrastructure, (5) testing practices and tooling, and (6) other practices and processes. Due to space limitations, we will briefly share some details about the lessons learned from just three of these categories. 


\emph{The core systems operated by the organization (14).} An incident can be a unique opportunity to see and learn about the behavior of one or more of the systems an organization owns, and specifically its behavior in a particular scenario. The result being a deeper understanding of a complex system and the space of possible behaviors. For example, during the postmortem of incident I12, the team learned that while their data storage approach ``has been advantageous for engineers'' it also has meant that ``a single change can have a far-reaching impact''. The learning process, in this case, is an opportunity to revisit design tradeoffs made.

\emph{Incident response practices and processes (12).} Postmortems involve a review of the incident response, providing an opportunity to learn (and improve) the processes and tools used. For example, though a deployment was temporally correlated with the start of incident I2, the responders dismissed the change and chose not to perform a rollback, and in retrospect decided that usually rolling back is the right decision, calling it ``less a lesson learned than a maxim ignored''. As part of responding to incident I11, the responders ``made several public estimates on time to repair'' that proved to be overly optimistic. They learned more about the variables that should be factored into such estimates.

\emph{Monitoring, logging and alarming systems (7).} These tools are important for both detecting and diagnosing failures. Broadly, the relevant lessons learned remind us that incident response and mitigation is helped by accurate, detailed, and timely information. Our data includes learnings around metrics information that was missing (e.g., providing ``limited visibility into job health'' (I5)) or unclear (e.g., ``drowned out by [...] unrelated exceptions coming in from other systems'' (I5)). During incident I12, the first indications of the failure came from (seemingly coincidental) downstream systems failures, and they learned that they were missing metrics showing the ``primary cause of the issue''.


\subsection{What Actions Come From Learning?}

In some of the incident reports we have analyzed, lessons learned are accompanied by (preventative or remedial) actions the organization plans to take in response to what was learned. Of the 50 lessons learned that we have identified in our preliminary analysis, 29 of them have at least one associated action and of those most have exactly one associated action, though some have as many as 4 actions. Note that in some cases, the associated action is a major undertaking and could easily have been decomposed into more actions, so the actual number of associated actions may not be too significant. For lessons in incident reports that do not have associated actions, it is possible that the actions are written down elsewhere, but in some cases, it seems a lesson is to informally inform future decision-making and actions. 

Actions can be seen as ways that the lessons learned can be acted on and potentially retained by the organization (beyond simply recording them in the incident report). They can also be seen as part of a larger process in which a socio-technical system is evolved based on learnings from failure analysis. We have found that actions may target core systems, procedures and processes, staffing, support systems, testing practices and more. Some actions aim to address a specific issue from the incident (e.g., fixing a particular defect) or something more general (e.g., addressing a class of similar potential defects). Example actions from our preliminary analysis include: 

\begin{enumerate}
    \item ``Add new load testing infrastructure'' to avoid deploying scalability regressions (I1),
    \item ``Fuzz older software looking for'' similar problems to the one featured in the incident (I4),
    \item ``Eliminate all unnecessary calls to the backend'' to improve efficiency (I7),
    \item Review and document configuration decisions to fill knowledge gaps (I12), and
    \item Analyze other ``database queries [...] which might pose similar risks'' (I13).
\end{enumerate}

We are continuing to analyze the actions that follow from lessons learned, with the goal of deepening our understanding of this failure-driven system evolution process. 

\subsection{Learning in Context}

Here we discuss several analytic concepts that provide important context for the learning process and provide insights into the role that lessons learned can play in evolving a socio-technical system. Note that most of these concepts emerged from our analysis of the interview data (though also in the incident reports to some extent). A more integrated analysis (based on both datasets) will be conducted in the future. 

\emph{Social construction of lessons learned.} Lessons learned are the product of a failure, but more specifically the product of a social process involving (disparate) interpretations and negotiation, as described by several of our participants (P2, P3, P4, P5, P7). A different set of individuals (or the same individuals at a different point in time) may arrive at a different set of conclusions, due to different perspectives, goals, and experience. In fact, at one organization, they hold ``multiple postmortems for an incident and based on the people [involved] we're going to discuss different stuff'' (P7).

\emph{Reminders and emphasis.} Some learnings are better understood as reminders of principles that the engineers already knew or previously learned. Some participants (P3, P7) found that incidents can help the team see the importance of certain principles, the potential risks if they are not followed, the context and scenarios in which they apply, and more. Also, repeated lessons learned are part of a reinforcement-learning like process.

\emph{Isolated learning.} We have seen little evidence of formal processes that involve learning across incidents and the purpose of incident reports is often to make sure a specific incident is not repeated: ``we can’t promise that we won’t go down again, but we can promise that we won’t make the same mistakes twice'' (I1). However, there are exceptions. For example, P2 found that ``sometimes you have to look at a group of [incident reports] to find the bigger trends''. After her organization had experienced multiple bad deployments (each covered in its own postmortem) she directed her engineers to look across all the incident reports and identify the ways their test environment was different from their production environment.

\emph{Situated lessons learned.} Related to the above, learnings are often strongly situated in a particular failure scenario, systems, and team context, limiting their transferability. P4 found that often his team identified lessons that were ``something that's extremely specific'' to the incident (such as ``we missed some tests'') but struggled to identify a ``meaningful learning that can be shared''. He also felt that some more experienced engineers were better at identifying something that can ``generalized'' and be ``disseminated''. 


\emph{Losing attention (learnings over time).} The outcomes of a postmortem, including the way that learnings are used, are not determined by the end of the postmortem (see also~\cite{haunschild2015organizational}). Instead, they unfold over time. Planned actions face organizational inertia, compete with previously planned roadmaps, and changing attention. P3 found that it was unreasonable to expect that everything planned during a postmortem would be completed. However, in the case of incident I7 (which was actually 6 separate incidents), the problems were so severe  that it completely shifted the mindset of the organization from innovation to safety. Improving the reliability of their system was prioritized ``above all else right now''. 

\emph{Learning amidst change and uncertainty.} Learning from failure is complicated by change and uncertainty in technology, context and environment, growth, and the system themselves~\cite{woods2018theory}. New lessons learned may make old learnings or planned actions obsolete. For example, a more abstract learning may lead to a new way to think about what actions are needed (see also~\cite{madsen2010failing}). As changes are made to systems (based on lessons learned) complexity may be added. P1 even found cases where changes based on lessons learned led to new incidents. Even after carefully analyzing a failure, it is still not necessarily obvious ``what the next thing is that's going to bite us'' (P3). This concept has implications for planning how to act on lessons-learned and suggests that lessons and plans may need to be revised overtime, as new learning opportunities are available. 
 
\emph{Recording lessons learned.} We have found that recording a lesson learned in an incident report does not mean that it has been learned by the organization. Similarly, analyzing an incident does not mean that an organization will be successful at preventing similar incidents in the future. Our participants described many challenges to disseminating lessons learned across an organization. To mention just one, P5 found that it was difficult to know which incident reports (created by the larger organization) had lessons that were pertinent to him and the systems he owns but did not have time to review all of them to find out (``maybe one every two or three weeks is worth reading'').


\section{Discussion}
\label{discussion}

As we gather more data and build on the preliminary findings just discussed, our goal is to deepen our understanding of the learning from failure process. Both formal and informal processes are of interest to us. We are also interested in understanding and helping alleviate the challenges organizations face in taking what was learned from a failure and successfully improving the reliability of the broader socio-technical system. Specifically, we have three main ambitions.

First, we intend to develop a model of failure-driven system evolution based on how systems fail, and the changes made in response. The model will account for a single failure and the changes proposed in response to that failure, but also larger failure patterns and changes proposed in response to multiple failures of a given system over time. Software evolution has received considerable attention in software engineering research, however, the social process of evolving systems based on failure has not yet been well studied.

Second, we will build an organizational learning model of the collaborative process that captures insights about how learnings and proposed changes are managed overtime. The organizational learning model will situate the failure-driven system evolution model just mentioned in an organizational context where there are costs, competing priorities, and employee turnover, and where risk and uncertainty play a role. How lessons learned are encoded (retained and transferred through the organization) is a key question.

We have found that lessons learned from incidents (including those described by our interviewees) vary along dimensions such level of abstraction, generality, and maturity. We suspect that lessons learned may often initially be narrow, applying only to a small part of a single system involved in the incident. How learnings change over time as more failures are experienced is an open question, though our findings suggest that they are generalized through informal processes and in some cases may contribute to the challenging process of learning from other's failures.

Third, we intend to develop and test interventions to improve aspects of the processes we have studied. The interventions will likely propose process changes and supporting tooling. The research methodology we have presented so far in this paper is largely oriented towards the first two ambitions, though we believe our emerging results have implications for the design of interventions, especially insights around how lessons learned are captured and shared. To further explore possible interventions, we plan to use prototyping and a participatory design methodology. Combining these methodologies will increase the likelihood that the interventions proposed will account for the complexity and context of real socio-technical systems.

\section{Conclusion}

We began this paper with a brief anecdote about a team of software engineers discussing a recurring lesson learned over lunch. After disagreeing about the lesson and expressing frustration about how often failures seemed to be caused by configuration changes, the team began discussing solutions. The lunch turned into an enthusiastic design session for a new type of tool (or service) for managing configuration changes, encoding the lesson learned in the tool so that it would be retained. Importantly the tool was to be generalized to be usable company-wide (facilitating transfer of the lesson learned). After the lunch one of the engineers wrote up the design and shared it with management with the hope that the work on the proposed tool might be funded. Five years later, work on the tool has not started, and we can only surmise that configuration changes continue to trigger incidents. 


\bibliographystyle{plain}
\bibliography{lfi}

\end{document}